\documentclass[proceedings]{ccn}
\ccndoi{10.32470/gn6tuko}  % required; assigned by CCN

\usepackage{graphicx}
\usepackage{amsmath}
\usepackage{amsfonts}
\usepackage{todonotes}

\newcommand{\EDIT}[1]{\textcolor{black}{#1}}

\NewDocumentCommand{\svd}{g}{%
\IfNoValueTF{#1}{\operatorname{svd}}%
{\,{\operatorname{svd}}\of{#1}}}

\title{Fast Whole-Brain, Geometry-Aware Functional Alignment for Cross-Subject Decoding}

\author{\centering
  Pierre-Louis Barbarant\affmark{1,2,3}
  \And Florent Meyniel\affmark{2, 3}
  \And Bertrand Thirion\affmark{1}
}
\affiliation{1}{Université Paris-Saclay, Inria, CEA, Palaiseau 91120, France}
\affiliation{2}{Cognitive Neuroimaging Unit, Institut National de la Santé et de la Recherche Médicale, Commissariat à l'Energie Atomique et aux énergies alternatives, Université Paris-Saclay, NeuroSpin center, 91191 Gif/Yvette, France}
\affiliation{3}{Institut de neuromodulation, GHU Paris, psychiatrie et neurosciences, centre hospitalier Sainte-Anne, pôle hospitalo-universitaire 15, Université Paris Cité, Paris, France}
\emails{pierre-louis.barbarant@inria.fr, bertrand.thirion@inria.fr}

\begin{document}

\maketitle

\begin{abstract}
    Decoding brain activity is useful for characterizing brain processes and understanding the functional architecture underlying cognition.
    However, the inter-individual variability in brain response patterns limits the development of decoders that generalize across individuals. 
    A solution to this challenge is functional alignment: aligning functional data across individuals before training population-level decoders. 
    The core issue is to strike the balance between aligning functional features and preserving the anatomical structure, while maintaining computational efficiency.
    We introduce a new functional alignment method for fMRI, SpectralOT, that embeds cortical geometry into Laplace-Beltrami eigenmodes along functional data to regularize the alignment.
\end{abstract}

%%%%%%%%%%%%%%%%%%%%%%%% INTRO %%%%%%%%%%%%%%%%%%%%%%%%
\section{Introduction}

Accurate brain activity decoders have now become ubiquitous in analyzing functional Magnetic Resonance Imaging (fMRI) data (\cite{Naselaris2011, Haynes2015, Liu2025ASO}).
The development of such decoders have enabled the study of increasingly complex cognitive processes by relating the activity of various brain regions to a rich set of stimuli.
However, the predictive power of population-level decoders trained through anatomically normalized data is hindered by the inter-individual variability of brain responses.
Building a robust population-level decoders requires to overcome not only the anatomical variability, but also the function variability between individuals.
More precisely, the two main issues related to inter-individual variability are: firstly,  the discrepancies in anatomy, yielding different spatial support for the functional signal, and secondly, the variability in the functional response location and magnitude (\cite{Haxby2001, Sabuncu2009}).
It is important to note that because both issues are entangled, solutions have to take both spatial and functional information into account.

In order to deal with anatomical variability across subjects, researchers have first developed and adopted accurate diffeomorphic matching to the MNI collection (\cite{Fonov2009}).
Surface-based registration has further improved alignment (\cite{Fischl1999, Robinson2014}).
However, these anatomical templates cannot match all details of the brain folding, and besides, functional architecture is not strictly tied to anatomical organization.
Given that the understanding of functional architecture involves accounting for spatial relationships (\cite{Lu2025}), the next frontier for the community is to create functional atlases that describe accurately the spatial organization of cognitive functions: functional templates, and the related registration procedures. While Optimal Transport (OT) solvers have shown promising performance in that respect, state-of-the-art approaches suffer from key limitations. Joint minimization of inhomogeneous criteria (functional matching, squared geodesic distance pairwise matching, marginal relationships of the coupling), makes solvers extremely costly and complex to run, and the relative weighting of these inhomogeneous terms remains a practical and computational hurdle. \EDIT{This computational burden precludes population-scale studies, nested cross-validation over hyperparameters, or integration with deep learning frameworks.}

In this work, we introduce SpectralOT, a method that relies on the \emph{functional maps} concept developed in the field of computer graphics. We adopt them with an OT loss tailored for registration. SpectralOT is much easier to parametrize, and the computational problem is orders of magnitude faster than current state of the art solution.

\subsection{Related Work}

\EDIT{All existing methods share the same pipeline: given paired fMRI recordings from two subjects, they estimate a vertex-to-vertex transformation that maps one subject's brain activity onto the other's anatomical space, then evaluate how well this transformation generalizes to held-out data.} We identified two distinct classes of functional alignment methods in the literature: local and whole-brain methods. 

\subsubsection{Local Methods} were introduced in the seminal work of \cite{Haxby2011} on Hyperalignment, where Procrustes transformations were derived between subjects using a searchlight (\cite{Kriegeskorte2006}), then aggregated into whole-brain transformations. The searchlight method ensures that global anatomical constraints are preserved by aligning data only locally, in a searchlight parcel. More recently, \cite{Bazeille2021} demonstrated that using non-overlapping parcels yield superior task decoding performance.

The main drawback of local methods stems from the selection of a searchlight radius or a parceled atlas, at the expense of introducing border effects which might harm spatial specificity of the signal. Moreover, these methods require spatial registration to a given anatomical template and cannot provide correspondences across individual-specific anatomical spaces (e.g. individual meshes).
We therefore focus on whole-brain alignment methods.

\subsubsection{The ProMises model} described in \cite{Andreella2022} strikes the functional-anatomical balance by incorporating an additive spatial prior into Procrustes decomposition of the cross-covariance of paired subject's signals. They derive an efficient low-rank implementation that scales Procrustes transform to tens of thousands of vertices and report increased between-subject classification accuracies on various tasks. \EDIT{Yet on limited data, the low-rank nature of the transformation undermines generalization to unseen functional patterns (see the first section of the Appendix).}

\subsubsection{Fused Unbalanced Gromov-Wasserstein (FUGW)} extended the OT approach to functional alignment by incorporating a quadratic Gromov-Wasserstein term that minimizes spatial distortion of transport maps (\cite{thual2022aligning}). This method increased Inter-Subject Correlation (ISC) (\cite{thual2022aligning}) and improved out-of-subject decoding performance in movie scenes retrieval (\cite{Thual2023AligningBF}). Unlike ProMises, FUGW generates full-rank transforms but requires tuning 3 distinct hyperparameters and lengthy GPU computation, which represents a major hurdle. In addition, FUGW balances inhomogeneous embeddings of functional (Wasserstein) and geometric information (Gromov-Wasserstein).

\subsubsection{Functional Maps} were developed in the field of computer graphics to deal with texture transfer across deformed meshes (\cite{Ovsjanikov2012}). By leveraging mesh-specific spectral descriptors, these methods construct correspondence matrices between spectral bases. However, these methods require the signal to be expressed in the spatial spectral domain. Preserving fine spatial patterns requires several hundred eigenmodes (\cite{Pang2023}) to achieve sufficiently short spatial wavelengths. Additionally, spectral descriptors struggle to capture the high-curvature geometry of the cortical surface (\cite{Abulnaga25}), limiting direct applicability of functional maps.

%%%%%%%%%%%%%%%%%%%%%%%% METHODS %%%%%%%%%%%%%%%%%%%%%%%%
\begin{figure}[!ht]
  \begin{center}
    \includegraphics[width=.85\linewidth, trim={1cm, 1.5cm, 1cm, 1.5cm}]{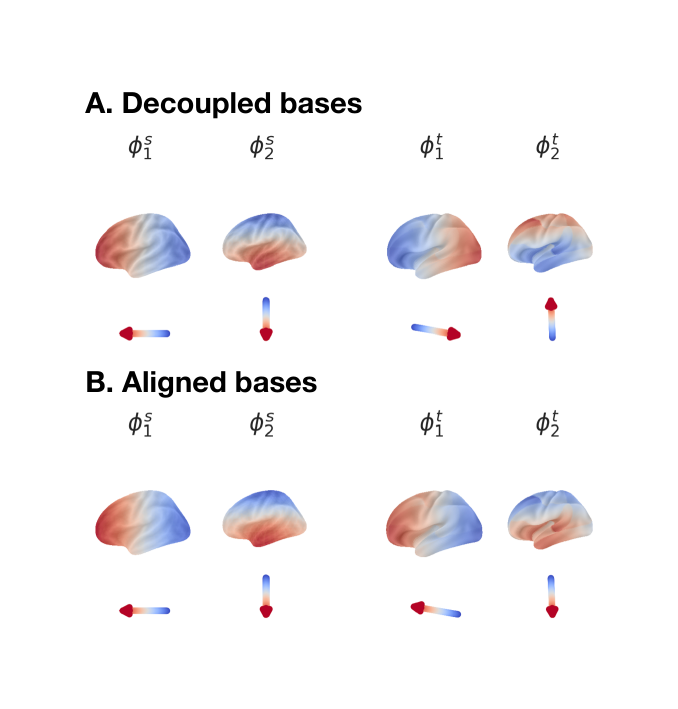}
  \end{center}
  \caption{
    \textbf{Aligning spectral bases across individuals.}
    By definition, eigenmodes of the Laplace-Beltrami operator are defined up to their sign. 
    \textbf{A.} This phenomenon may lead to a "sign flip", where corresponding eigenmodes inconsistently describe intrinsic geometries across meshes.
    \textbf{B.} We rely on a simple procedure to derive aligned bases. For each pair of eigenmodes, we compute the gradient of values in each mesh 3D coordinate space. This gradient is represented in the Figure by a colored arrow under each mesh.
    Leveraging the common orientation between spaces, we flip the sign of an eigenmode if the dot product between gradients is negative.
    Repeating this procedure across every pairs of eigenmodes yields aligned descriptions of geometry across meshes.
  }
  \label{fig:sign-flip}
\end{figure}

\section{Methods}

We seek to develop an effective method where the correspondence between the peaks of a source subject and a target subject is determined locally by signal similarity, while preventing the pairing of regions located in distant brain areas.
Given two cortical meshes with $n$ and $m$ vertices, we denote by $F^s \in \mathbb{R}^{p\times n}$ and $F^t \in \mathbb{R}^{p\times m}$ the functional signal in the source and target subjects, respectively, for a set of $p$ paired stimuli. Our goal is to obtain a correspondence matrix $\mathbf{P}^{s \rightarrow t} \in \mathbb{R}^{n\times m}$, a.k.a \emph{coupling}, between the source and target meshes that can be used to transfer new signal across individuals.

\subsection{Geometric Descriptors}
To prevent distant vertex correspondences, we construct a dissimilarity matrix $C_{geom} \in\mathbb{R}^{n\times m}$ penalizing far-apart vertices. Given that the shape of the source and target meshes may differ in shape and resolution, our goal is to find intrinsic descriptors of the anatomical structure that are invariant under scaling, translations, rotations (near identity), resampling, or small deformations.

Inspired by the work of \cite{Rustamov07}, we characterize a vertex $v$ by the set of Laplace-Beltrami eigenmode values $\{\phi_1(v), \phi_2(v), \cdots\}$, obtained by diagonalizing the Laplace-Beltrami Operator (LBO):
\begin{align}
    -\Delta \phi = \lambda \phi
\end{align}
where $-\Delta$ is the (negative) mesh Laplacian and $\lambda$ the eigenvalue associated to $\phi$. The set of weighted eigenmodes $\left(\frac{\phi_1}{\sqrt{\lambda_1}}, \frac{\phi_2}{\sqrt{\lambda_2}}, ...\right)$ forms an orthonormal basis on the mesh invariant under isometric (distance-preserving) transformations (\cite{Rustamov07}), creating a robust embedding of mesh geometry in the functional domain. Given approximately isometric brain meshes, we construct a vertex-to-vertex dissimilarity matrix $C_{geom} = ||\phi^s - \phi^t||^2_2 $ encoding spectral embedding distances between source vertex $i$ and target vertex $j$. We illustrate this step in panel A. of Figure \ref{fig:overview}.

In practice, the dissimilarity matrix $C_{geom}$ penalizes long-distance correspondence because it is computed using only the low-frequency eigenmodes corresponding to the largest wavelength. We truncate the bases to the first three eigenmodes, which encode anterior-posterior, dorsal-ventral, and lateral directions respectively (Figure \ref{fig:overview}, panel A). \EDIT{Sensitivity analyzes show minimal improvement above three components (see Appendix Figure \ref{fig:eigenmodes-appendix}).} Eigenmodes are computed using the \verb|robust_laplacian|\footnote{\url{https://github.com/nmwsharp/robust-laplacians-py}} Python package (\cite{Sharp:2020:LNT}). The eigenvector associated with the null eigenvalue is excluded from the analysis.

\subsubsection{Sign-flips} arise because of the eigenmodes sign indeterminacy \cite{Kovnatsky2013}: if $\phi$ is an eigenmode then $- \phi$ is also an eigenmode (Figure \ref{fig:sign-flip}). To obtain aligned bases across meshes, we use a simple scheme: for each pair of eigenmodes across meshes, we compute 3D gradients of eigenmodes values treating meshes as point clouds, then evaluate the gradients dot product sign. If the sign is negative, we flip the sign of one eigenmode. This scheme is valid as long as meshes orientations are similar, a condition typically satisfied in fMRI where subjects are registered to common coordinate spaces.

\subsection{SpectralOT}

\begin{figure}[!ht]
  \begin{center}
    \includegraphics[width=.85\linewidth, trim={1.3cm 1.3cm 1cm 1.5cm}, clip]{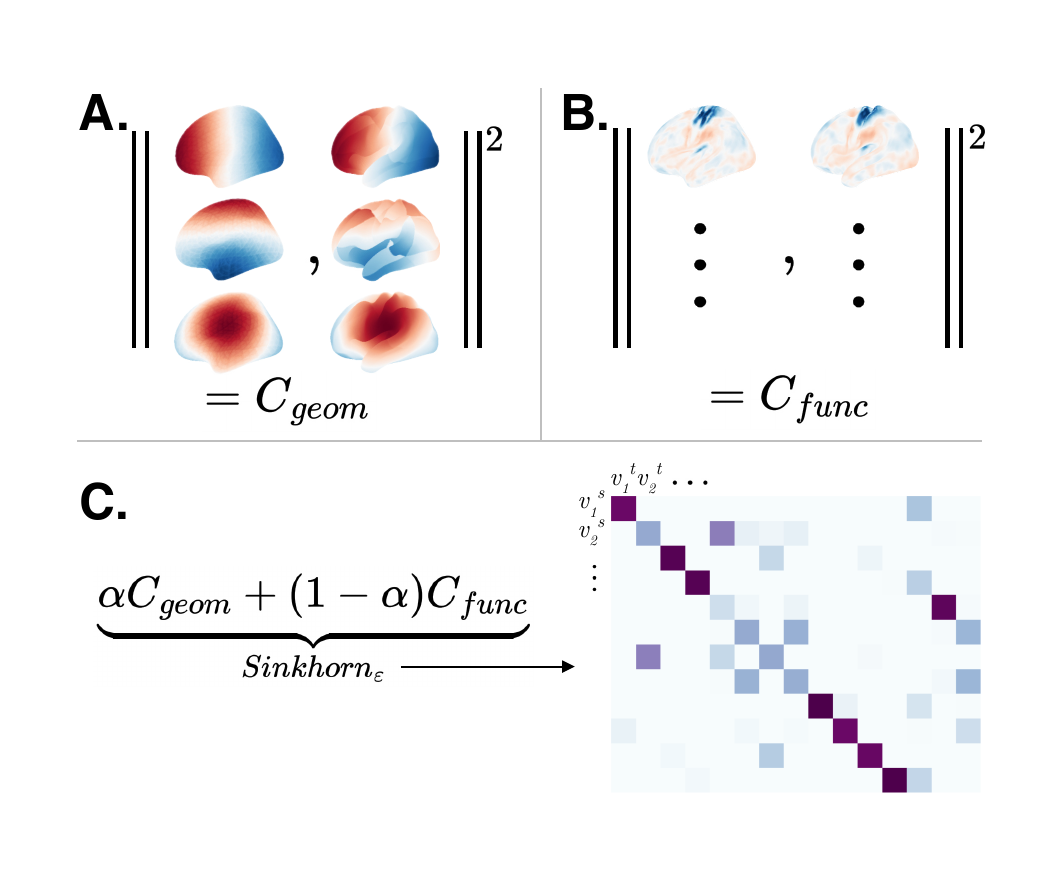}
  \end{center}
  \caption{
    \textbf{Overview of the SpectralOT method.}
    Our procedure consists of three distinct steps. For each hemisphere:
    \textbf{A.} We first compute the first three eigenmodes of the Laplace-Beltrami operator on both source and target mesh and compute the vertex-wise squared $L^2$ distance across the eigenmodes to derive the geometric cost matrix $C_{geom}$.
    \textbf{B.} We then compute the vertex-wise squared $L^2 $ distance between the source and target functional data across paired samples to derive the functional cost matrix $C_{func}$.
    \textbf{C.} We proceed by interpolating between both cost matrices using a hyperparameter $\alpha$ that controls the respective strength of each component. This composite cost is then passed to a Sinkhorn solver to obtain a soft correspondence matrix $\mathbf{P}^{s\rightarrow t}$ between source and target vertices.
  }
  \label{fig:overview}
\end{figure}

To incorporate functional information to our model, we compute a vertex-wise dissimilarity matrix $C_{func} = ||F^s-F^t||^2_2 \in \mathbb{R}^{n\times m}$ across paired samples of source and target data (Panel B of Fig. \ref{fig:overview}).
Putting our results together, we obtain a composite cost $C_{composite} = (1-\alpha ) C_{func} +  \alpha C_{geom}$ that interpolates functional and geometric constraints using a single parameter $\alpha \in [0,1])$. Note that this composite cost characterize whole-brain data. Following the approach of \cite{Pai_2021_CVPR}, we pass this cost to a Sinkhorn Entropic OT solver (\cite{Cuturi13}) to obtain a coupling $\mathbf{P}^{s\rightarrow t} \in \mathbb{R}^{n\times m}$ between source and target vertices (Panel C of Fig.\ref{fig:overview}). Specifically, $\mathbf{P}^{s\rightarrow t}$ solves the entropic optimal transport problem (\cite{peyre-cuturi}): 
\begin{align}
    \min_{\mathbf{P} \in \mathbb{R}^{n\times m}} \text{Tr}(\mathbf{P} \mathbf{C}) &+ \varepsilon \mathbf{E}(\mathbf{P})  \\
\text{s.t.} \quad \mathbf{P} \mathbf{1}_m = \frac{\mathbf{1}_n}{n} \ \text{and}& \ \mathbf{P}^T \mathbf{1}_n = \frac{\mathbf{1}_m}{m}  
\end{align}
where $\varepsilon > 0$ controls the entropic penalty:
\begin{align}
    \mathbf{E}(\mathbf{P}) = \sum_{i=1}^n \sum_{j=1}^m \mathbf{P}_{i,j} \left( \log (\mathbf{P}_{i,j}) - 1 \right)
\end{align}

In practice, computation of the couplings are carried out using \texttt{PythonOT}'s log-Sinkhorn solver (\cite{flamary2021pot}). We divide both costs $C_{func}$ and $C_{geom}$ by their respective maximum value to equally balance each term's contribution to the composite cost and avoid numerical overflow. \EDIT{The entropic regularization term is different in FUGW and SpectralOT (quadratic and linear,  respectively); we set $\varepsilon = 10^{-4}$ for FUGW and $\varepsilon = 10^{-3}$ for SpectralOT unless otherwise specified. These values yield comparable blurring of the transport plans (see Appendix Figure~\ref{fig:epsilon-appendix}).}
%Unless otherwise specified, we set the entropic regularization strength $\varepsilon$ to $10^{-3}$, the smallest value ensuring Sinkhorn convergence under 1000 iterations.
%
We set $\alpha$ based on the desired anatomical regularization strength. The edge case $\alpha =0$ completely lifts the geometric constraint and $\alpha =1$ drives the alignment purely on anatomical features. As in \cite{thual2022aligning}, this parameter can be set via grid-search.

Given new data $F^{test}$ on source subjects vertices, we generate predictions leveraging the computed coupling:
\begin{align}
    \widehat{F^{test}} = \left(\frac{
        \sum_{i=1}^n \mathbf{P}_{i,j}^{s\rightarrow t} F^{test}_{i} 
    }{
        \sum_{i=1}^n \mathbf{P}_{i,j}^{s\rightarrow t}
    }\right)_{j=1}^m \label{eq:transform}
\end{align}

Initializing the solver with equal mass across vertices yields marginal condition $\sum_{i=1}^n \mathbf{P}_{i,j}^{s\rightarrow t} = 1/m$ for every source vertex $i$, simplifying equation (\ref{eq:transform}) to $\widehat{F^{test}} = m\mathbf{P}^{s\rightarrow t} F^{test}$.

Compared to FUGW (\cite{thual2022aligning}), which requires multiple Sinkhorn solver runs, SpectralOT requires only one pass, drastically reducing the complexity of the matching scheme while maintaining an anatomical component.

\subsection{Experiments}

We assess SpectralOT correspondence quality through three experiments of increasing complexity, comparing against state-of-the-art FUGW and ProMises models with anatomical registration as baseline.

\subsubsection{Experiment 1: Anatomical Alignment} We align \EDIT{both} hemispheres of \textit{fsaverage5} (\cite{fsaverage}) and \textit{fsLR} (\cite{VanEssen2011}) surface templates using solely anatomical information, comparing how well shape descriptors encode geometry versus pairs of euclidean distances used in the Gromov-Wasserstein (GW) component of FUGW.
We perform color transfer of the \textit{Destrieux} surface atlas (\cite{Destrieux2009}), visualizing the mapping of distinct parcels across meshes. Since a row-normalized transport plan yields a probability distribution of being associated with every source vertex for each target vertex, we compute the negative entropy to measure probability mass spread.
Ensuring fair comparisons, we set the number of Sinkhorn iterations to 1,000 and the entropy regularization parameter $\varepsilon$ to $10^{-4}$. Additionally for FUGW, we set hard marginal constraints ($\rho=+\infty$) to ensure complete probability mass transport and provide 10 iterations of block coordinates following \cite{Thual2023AligningBF}.
The ProMises model is excluded from this experiment as it requires identical source and target meshes.

\subsubsection{Experiment 2: Task-Based Functional Alignment} We perform functional alignment between \EDIT{every} pair of the IBC dataset (\cite{Pinho2024}) with varying $\alpha$ levels. We rely on fixed-effect contrast maps computed on the \textit{fsaverage5} surface template.
40 paired samples from \textit{ArchiStandard, ArchiSpatial, ArchiSocial}, and \textit{ArchiEmotional} task batteries build the alignment set, while 16 paired \textit{MathLanguage} contrast maps are used to assess alignment quality via inter-subject Pearson correlation (ISC) after source-to-target projection (\cite{thual2022aligning}).
We report total computation time for both hemispheres for FUGW, ProMises, and our method, setting $\varepsilon$ to $10^{-4}$ for FUGW following \cite{Thual2023AligningBF} and $10^{-3}$ for SpectralOT. This discrepancy arises from the rescaling of FUGW's "fused" cost when passed to the Sinkhorn solver (see algorithm 3 of \cite{thual2022aligning}).

\subsubsection{Experiment 3: Cross-Subject Decoding} ISCs are inherently biased by the smoothing effect of entropic regularization (\cite{Bazeille2019}). In addition, we used cross-subject decoding, which is less sensitive to smoothing. We design a decoding experiment using the \EDIT{three publicly available subjects from the} Courtois Neuromod THINGS (\cite{St-Laurent2025}) dataset.
Our goal is to evaluate how effectively the proposed method reduces domain shift across subjects. To do this, we compute alignments for all possible subject pairings. We use a subset of images shared across all participants to estimate functional alignments, and then train a classifier over each target subject using held-out data.
We evaluate the accuracy of this classifier on aligned source subject data. Decoders classify 27 object categories associated to the THINGS stimuli (\cite{Hebart2019}) using activation maps from three publicly available participants of the study.
We rely on \texttt{scikit-learn}'s (\cite{scikit-learn}) \texttt{LinearSVC} implementation for classification. High quality transformations should improve decoding over anatomical registration on \textit{fsaverage5}.
%, which is used to derive baseline scores.
\EDIT{\subsubsection{Experiment 4: Group-Level Decoding} The previous dataset encompasses only three densely-sampled subjects, limiting statistical
comparisons between methods, and focuses exclusively on naturalistic stimuli.
We therefore design a group-level decoding experiment on the IBC dataset (13 participants),
inspired by the framework of~\cite{Bazeille2021}, to more broadly assess the performance of
SpectralOT against other methods.
We use single-trial z-scored condition maps from the \emph{RSVPLanguage} protocol (360
samples per subject). For each subject, data are split into three folds (33\% alignment, 66\% decoding), stratified by run. Within each fold, we perform leave-one-subject-out cross-validation: the held-out subject serves as the common alignment target, and the remaining subjects' data are aligned to it using the corresponding alignment split. A \texttt{LinearSVC} is then trained on the aligned decoding data from the remaining subjects and evaluated on the held-out subject's decoding data, yielding a single accuracy score per fold.
}

%%%%%%%%%%%%%%%%%%%%%%%% RESULTS %%%%%%%%%%%%%%%%%%%%%%%%
\section{Results}

\begin{figure*}[!ht]
  \begin{center}
    \includegraphics[width=.9\textwidth, trim={0cm, 0.5cm, 0cm, 1cm}]{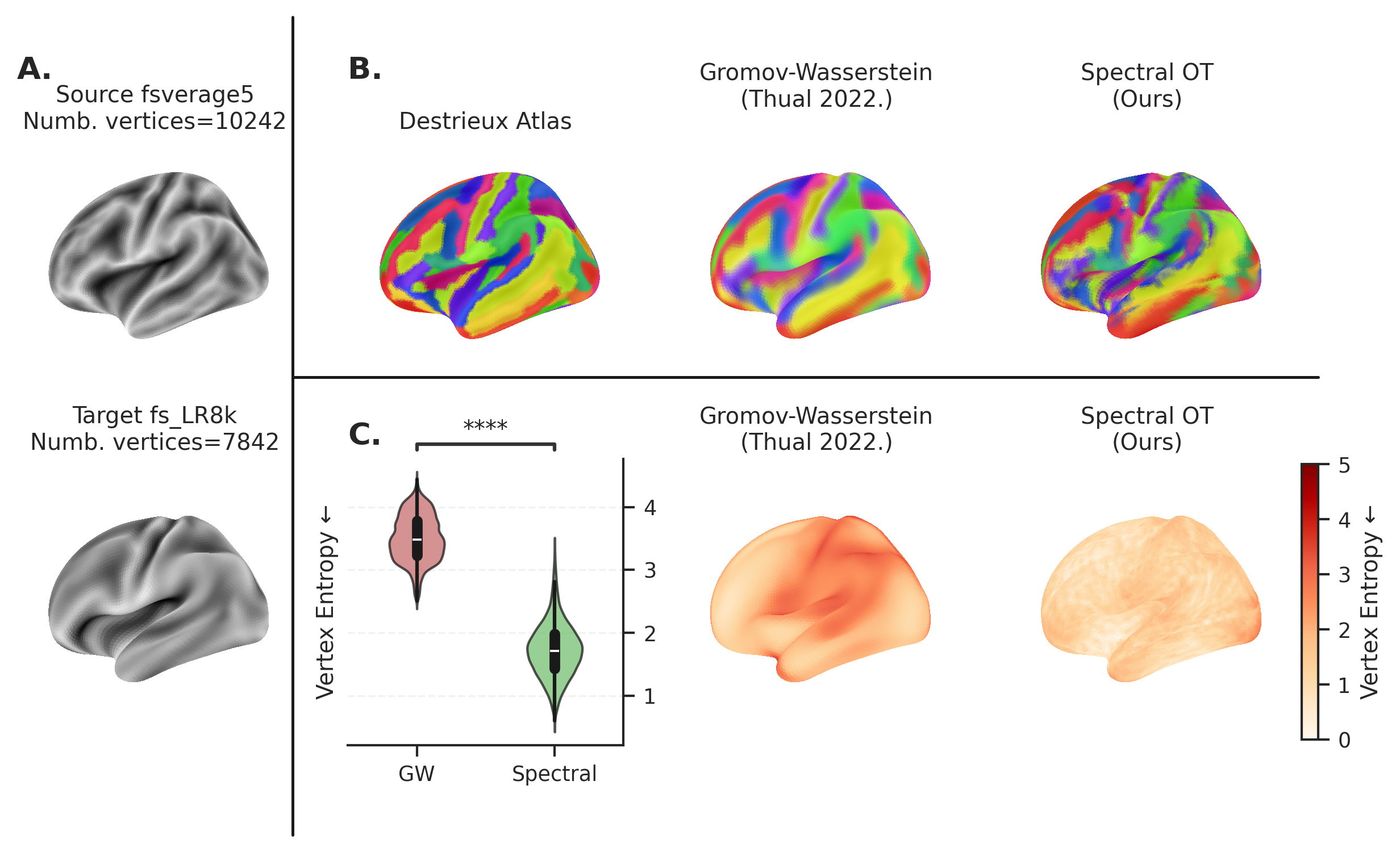}
  \end{center}
  \caption{
        \textbf{Anatomical alignment across different meshes ($\alpha=1$).}
        We compare our method to Gromov-Wasserstein (GW) solvers in their ability to derive purely anatomical correspondence between two different meshes.
        We set both OT solvers entropic regularization $\varepsilon$ to $10^{-4}$, with 1000 iterations of Sinkhorn's algorithm and 10 steps of block-coordinate descent for GW.
        \textbf{A.} We use left hemispheres of \emph{fsaverage5} and \emph{fsLR} surface templates as source and target respectively.
        \textbf{B.} Color transfer of the \emph{Destrieux} atlas between meshes using the computed transformations across vertices. Color blending is performed in RGB space to showcase information mixing across distinct areas of the atlas.
        \textbf{C.} Normalizing a transport plan across rows yields for each target vertex the probability of being associated with each of the source vertex. Typically, most of the incoming probability mass should come from a few number of vertices. We quantify this property by measuring the entropy associated to each target vertex's probability distribution (lower is better).
    }
  \label{fig:anatomical-example}
\end{figure*}

We first examine the ability of our method to produce anatomically coherent transformations compared to FUGW (Figure \ref{fig:anatomical-example}). We then incorporate signal from IBC participants to study the effect of interpolation between functional and anatomical constraints on ISC (Figure \ref{fig:isc}). Finally, we perform cross-subject \EDIT{and group-level} decoding to assess domain-shift reduction across individuals (Table \ref{tab:accuracies}, \EDIT{Figure \ref{fig:ibc-decoding}}).

\subsection{Softer Anatomical Priors}

Figure \ref{fig:anatomical-example} showcases alignment between the \textit{fsaverage5} and \textit{fsLR} inflated left hemispheres surface templates using only anatomical information. Panel B demonstrates that despite using only the first three LBO eigenmodes encoding the principal mesh direction, SpectralOT correctly transport the \textit{Destrieux} atlas across meshes at whole-brain scale. In particular, SpectralOT does not shuffle the different parcels across the cortex, indicating preserved global properties of brain geometry after transformation. \EDIT{Alignment of the right hemispheres (Appendix Figure \ref{fig:anatomical-example-appendix}) leads to similar conclusions.}

Compared to FUGW, SpectralOT exhibits less fine-scale regularity, reflecting the low sensitivity to local geometric differences captured in Laplace-Beltrami eigenmodes.
At equal levels of entropic regularization ($\varepsilon=10^{-4}$), FUGW generates blurrier signal on the target mesh, while SpectralOT's transported atlas shows less spatial smoothing. \EDIT{This trend becomes more pronounced at higher values of $\varepsilon$, as expected from
the different nature of entropic regularization in each method
(see Appendix Figure~\ref{fig:epsilon-appendix}).}

\subsection{Parsimonious Correspondence}

Panel C of Figure \ref{fig:anatomical-example} shows that SpectralOT concentrates the probability mass on a smaller set of vertices than FUGW, indicating higher parsimony in vertex-level associations. FUGW associates a larger number of source vertices to target vertices located along folds of the target mesh, revealing FUGW's transport plan dependence on sulcal curvature that increases signal mixing around high-curvature vertices.

\subsection{Enhanced Inter-Subject Correlation}

\begin{figure*}[!ht]
  \begin{center}
    \includegraphics[width=.8\textwidth, trim={1cm, 0.5cm, 1cm, 1cm}]{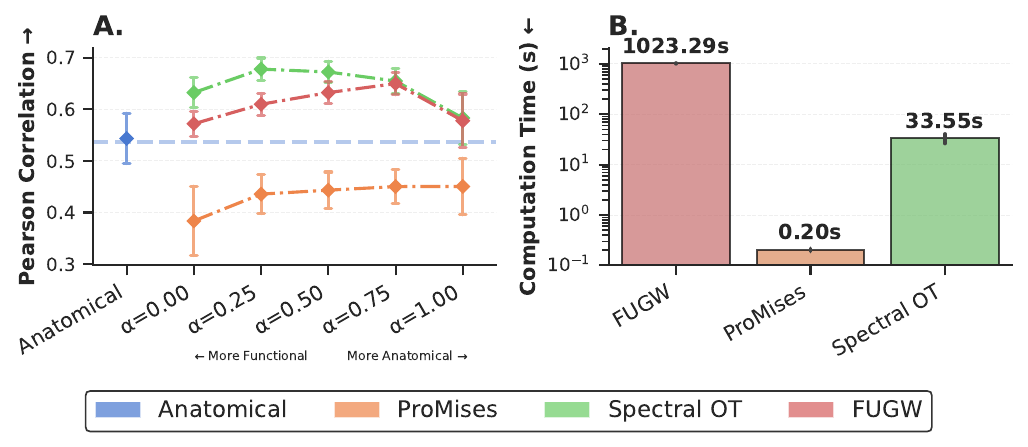}
  \end{center}
  \caption{
    \textbf{Inter-subject alignment on task data.}
    We align IBC dataset's subject $\#4$ on subject $\#9$. We use paired samples from the \textit{Archi} task battery to derive alignment transformation.
    We map left-out testing data from the \textit{MathLanguage} protocol between subjects and compare the aligned data to the target's subject test data.
    \textbf{A.} We report the Pearson correlation between each pair of samples in the aligned and target subjects data.
    We compare our method against FUGW and ProMises models for varying levels of anatomical regularization (higher $\alpha$ associates a stronger weight to the spatial component).
    Anatomical registration on \EDIT{both \textit{fsaverage5} hemispheres} is used to derive baseline scores.
    \textbf{B.} We report the average time taken by each method to process data on both hemispheres across the various regularization levels.
    We consider the total time taken to compute the alignment map as well as transforming the test data.
    All methods run on similar hardware to ensure fair comparisons.
    We set entropic regularization parameter $\varepsilon$ to $10^{-4}$ and \EDIT{$10^{-3}$} for FUGW and our method respectively.
  }
  \label{fig:isc}
\end{figure*}

Panel A shows SpectralOT surpassing the anatomical baseline for every value of geometric regularization $\alpha$, but also surpassing both FUGW and the ProMises model in Inter-Subject Pearson Correlation (ISC). Notably, the highest average ISCs occur for $0< \alpha <1$, suggesting that the inclusion of both anatomical and functional constraints leads to improved functional alignments, \EDIT{a trend that holds consistently across all subjects
(see Appendix Figure~\ref{fig:isc-averaged-appendix}).} When using purely anatomical alignment ($\alpha =1$), both FUGW and SpectralOT outperform the baseline. We attribute this observation to the inherent spatial smoothing related to entropic regularization in OT (\cite{progressive-transport}) which increases the ISC (\cite{Bazeille2019}).

Finally, it appears that the Promises model falls short compared to the other methods \EDIT{on all pairs of subjects (see Appendix Figure \ref{fig:isc-complete-appendix})}, including the anatomical baseline. The issue actually stems from the mathematical specification of the model: 
The linear model is limited to the span of the data used for alignment (see the Appendix), hence, even when no data reduction is performed
(\cite{Andreella2022}), we obtain a highly low-rank mapping between subjects vertices. Such mapping fails to transmit information on previously unseen data, including in the purely anatomical case $\alpha=1$ as the geometric prior is compressed.

\subsection{Balanced Constraints}

\begin{table*}[!ht]
  \begin{center}
    \caption{
        \textbf{Across-subjects decoding}
        For each pair of source/target subjects in the \emph{THINGS} dataset we compute an alignment map based on paired dedicated samples.
        We train a classifier on the other trials of the target subject to decode the 27 higher-level categories of objects.
        Left-out-data from the source subject are aligned and pass it to the trained classifier to derive a single accuracy score.
        Tighter alignment between source and target subjects results in better scores.
        We use Anatomical registration on the \textit{fsaverage5} template to compute baseline scores.
        We set the spatial regularization $\alpha$ to 0.5 for all methods. Entropic regularization $\varepsilon$ is set to $10^{-4}$ and $10^{-3}$ for FUGW and our method respectively.
        Bold results indicate highest score and underlined second best score.
    }
    \label{tab:accuracies}
    \vskip 0.12in
    \begin{tabular}{r|cccc}
      \hline
       Method& Anatomical & FUGW & ProMises & SpectralOT \\
       Source $\rightarrow$ Target & (Baseline)& (\cite{thual2022aligning}) & (\cite{Andreella2022})& (Ours)\\
        \hline\hline
            sub-01 $\rightarrow$ sub-02 & \underline{0.128} & 0.114 & 0.078 & \textbf{0.150} \\
            sub-01 $\rightarrow$ sub-03 & 0.080 & \underline{0.100} & 0.058 & \textbf{0.135} \\
            sub-02 $\rightarrow$ sub-01 & \textbf{0.120} & 0.083 & 0.073 & \underline{0.112} \\
            sub-02 $\rightarrow$ sub-03 & \underline{0.087} & 0.064 & 0.056 & \textbf{0.088} \\
            sub-03 $\rightarrow$ sub-01 & \underline{0.170} & 0.126 & 0.078 & \textbf{0.191} \\
            sub-03 $\rightarrow$ sub-02 & 0.140 & \textbf{0.163} & 0.109 & \underline{0.162} \\ \hline
            Average score               & \underline{0.121} & 0.108 & 0.075 & \textbf{0.140} \\
            Standard deviation          & 0.031 & 0.032 & 0.017 & 0.033 \\
      \hline
    \end{tabular}
  \end{center}
\end{table*}

The ISC curves profile on Panel A of Figure \ref{fig:isc} demonstrates strong skewness towards higher $\alpha$ values for FUGW, whereas SpectralOT exhibits a more balanced profile.
This illustrates a fundamental difference in how anatomical components are handled. SpectralOT linearly interpolates both the functional and geometric costs, resulting in balanced losses. In contrast, FUGW’s functional and geometric components scale as $\mathcal{O}(n^2)$ and $\mathcal{O}(n^4)$ respectively, where $n$ is the number of vertices) (\cite{thual2022aligning}).
This imbalance results in a nonlinear influence on $\alpha$, which affects the skewness of the ISC profile.

\subsection{Computational Efficiency}

We report in Panel B of Figure \ref{fig:isc} the total computation time (encompassing alignment and projection of test data from the source to the target subject on both hemispheres).
All algorithms have been implemented using \textit{PyTorch} (\cite{pytorch}) and run on an NVIDIA GeForce GTX 1080 Ti GPU. We recall that for both SpectralOT and FUGW the number of iterations of Sinkhorn has been set to 1000, with 10 iterations of block-coordinate descent for FUGW.
SpectralOT requires only \EDIT{33.55s, 30.5× faster than FUGW (1023.29s). ProMises is fastest (0.2s)} but delivers poor ISC performance due to low-rank data representation.

\begin{figure}[!ht]
  \begin{center}
    \includegraphics[width=.85\linewidth, trim={1cm, 0.5cm, 0cm, 1cm}]{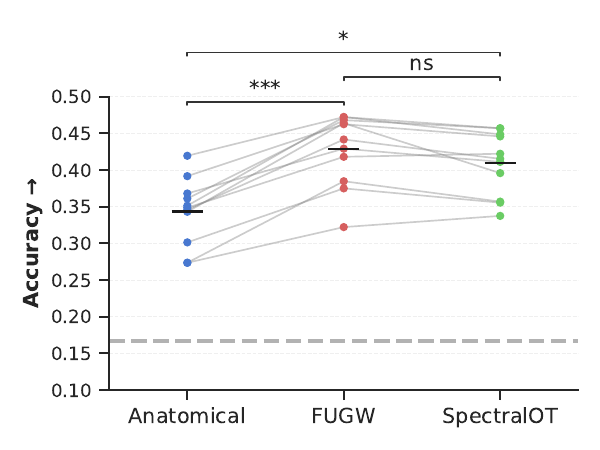}
  \end{center}
    \caption{\EDIT{
        \textbf{Decoding performance on the IBC dataset.}
        Inter-subject decoding accuracy for FUGW, SpectralOT, and the anatomical baseline.
        Each colored dot represents classifier accuracy for a left-out subject, averaged
        across all cross-validation folds. Solid lines connect subjects across methods.
        The dotted gray line indicates chance level.
        Statistical comparisons between all method pairs are performed using two-sample
        $t$-tests with \cite{Nadeau99} variance correction, under the null hypothesis of
        no difference in performance.
        The number of block-coordinate descent iterations of FUGW are reduced to $5$ due to the large number of alignments.
        Significance levels: \textit{*} $p\leq 0.05$, \textit{**} $p\leq 0.01$, \textit{***} $p\leq 0.001$.
    }}
  \label{fig:ibc-decoding}
\end{figure}

\subsection{Reduction of Domain Shifts}

While SpectralOT surpasses other methods in terms of ISC, we note that ISCs are biased by smoothing (e.g. the entropic term in the coupling estimation). To assess how well SpectralOT bridges the domain shift across individuals, a third experiment measures out-of-subject generalization (\cite{Thual2023AligningBF}) on naturalistic data. We rely on data from three participants from the Courtois-Neuromod THINGS dataset (\cite{St-Laurent2025}).
Out-of-generalization scores are reported in Table \ref{tab:accuracies} for various methods and combinations of source/target subjects. We observe that SpectralOT surpasses \EDIT{the anatomical baseline in five out of six pairs of source and target subjects, more than any other method. }

\EDIT{To statistically assess the improvement yielded by our method, we perform a group-level decoding experiment (experiment~4) on the 13 subjects of the IBC dataset, using task data from the \emph{RSVPLanguage} protocol. ProMises is excluded from this comparison due to its low performance in experiments~2 and~3. Both FUGW and SpectralOT achieve statistically higher decoding accuracy than the anatomical baseline (Figure \ref{fig:ibc-decoding}, $p \leq 0.05$), with no significant difference between the two methods.}

%%%%%%%%%%%%%%%%%%%%%%%% DISCUSSION %%%%%%%%%%%%%%%%%%%%%%%%
\section{Discussion}

We introduced SpectralOT, a functional alignment methods that can operate with different individual meshes to bridge inter-subject brain imaging-domain differences. SpectralOT builds an anatomical prior by leveraging spectral embeddings of the geometry of the cortical surface, and generates a composite cost with the dissimilarity matrix of the functional signal. This composite cost is then used for soft-matching vertices in a source and target subject, with an entropic OT solver. Our experiments show that our approach correctly captures the global geometry of the cortex while creating more parsimonious correspondence than the state-of-the art FUGW method. We also demonstrate higher cross-subject correlations, as well as increased performance in out-of-subject decoding. Finally, we highlight that our method converged several orders of magnitude faster than FUGW. In the following section, we outline the main implications of this novel method.

\subsection{Narrowing the Subject Shift}

\EDIT{Experiments~2 to~4 confirm that SpectralOT consistently outperforms both the anatomical baseline and ProMises, while yielding decoding performance statistically equivalent to FUGW. These results are consistent across diverse datasets spanning task-based and naturalistic paradigms. Notably, SpectralOT yields the fewest decoding scores below the anatomical baseline of all the comparisons. These results establish SpectralOT as a safe and robust candidate for broader applications.}

\subsection{Data Frugality}

Our analyzes revealed that while frameworks such as the ProMises model required a large amount of data to generate useful functional alignments, SpectralOT, similarly to FUGW, is able to bridge across-subjects domain shifts using only tens of localizer contrast maps. Since functional alignment inevitably requires a set of paired left-out data for both source and target subjects, we emphasize that the frugal aspect of our method makes it convenient for smaller-scale studies.

\subsection{Practical Advantages}

SpectralOT's rapid convergence compared to FUGW makes it appealing for fast iterations of experiments. Additionally, the linear dependence of the composite cost on the $\alpha$ parameter makes our method easier to tune and interpret, whereas FUGW's non-linear dependence on $\alpha$ severely hinders the possibility to set it with nested cross-validation. These aspects make our method the preferred choice especially as the number of alignments to fit grows with the number of subjects.

\subsection{Limitations and Future Directions}

Having established the superiority of our method in pairwise decoding, it would be interesting to assess its capacity in generating functional templates. Developing a template framework as originally done for Hyperalignment (\cite{Haxby2011, Guntupalli2016}) would yield the benefits of having a single common functional space aggregating the information of an entire cohort of subjects. Functional templates can be used in particular as a common target to align new data, reducing the complexity of experiments by avoiding computing alignments between every combination of subjects (\cite{Barbarant25}).

Next steps would include extending our framework to brain volumes. Since we rely on robust estimation of the LBO eigenmodes that supports both mesh and volume data, deriving a cost matrix associated to the volume geometry is straightforward.
\EDIT{A more exhaustive comparison would further benefit from 
data-driven selection of $\alpha$ and $\varepsilon$ 
across a broader range of datasets.}
Finally, we underline that computations of the transport plan based on entropic OT are differentiable with respect to every component of the composite cost (\cite{Cuturi19}). More specifically, SpectralOT can be integrated among neural network modules, which would be of great interest for studies leveraging deep-learning.

\subsection{Conclusion}

Overall, SpectralOT advances towards reducing inter-individual domain shifts, improving the performance in out-of-subject generalization of decoders. Our method is fast, easy to tune and generates parsimonious alignments of higher quality than state-of-the art approaches. We share our implementation with the community at the following link: \url{https://github.com/pbarbarant/spectralot}.

\clearpage

\subsection{Acknowledgments}
This work has benefited from State support as part of the Audace! Programme led by the CEA and managed by the Agence Nationale de la Recherche under the France 2030 heading, bearing the reference ANR-24-RRII-0004. This project was supported as part of the DANDI Inria-UdM associate team project, with support from Agence Nationale de la Recherche (ANR-25-CE23-3590), NSERC (RGPIN-2025-06022), and IVADO (602729-2024). Usage of LLM tools was strictly limited to rephrasing and improving the language.

\printbibliography

\appendix

\onecolumn
\section{Appendix}

\subsection{Procrustes Alignment}
Recall our notations: we denote $n$ the number of vertices of a mesh and $p$ the number of paired samples across individuals. The classical Procrustes model described by \cite{Haxby2011} computes an orthogonal transformation (rotation or reflection) $R$ between matrices of activity $F^s \in \mathbb{R}^{p\times n}$  and $F^t \in \mathbb{R}^{p\times n}$ of a source and target subject:
\begin{align}
    U, \Sigma, V^T &= \svd\left((F^s)^T F^t\right) \label{svd-procrustes}\\ 
    R &= UV^T
\end{align}
where $\svd$ denotes the Singular Value Decomposition (SVD) of a matrix.

\subsection{ProMises Model}
\cite{Andreella2022} proposes to add a spatial prior of the form $k\exp{(-D)}$ where $D$ is the matrix of euclidean distances between vertices on a similar mesh and $k\in\mathbb{R}$ controls the regularization strength. This term is incorporated into the SVD in equation (\ref{svd-procrustes}):
\begin{align}
    U, \Sigma, V^T &= \svd\left((F^s)^T F^t + k\exp{(-D)}\right) \label{svd-promises}
\end{align}

To make comparisons with SpectralOT and FUGW, we rewrite the inside term of the SVD to interpolate between both functional and anatomical component, equation (\ref{svd-promises}) thus becomes:
\begin{align}
    U, \Sigma, V^T &= \svd\left((1-\alpha)(F^s)^T F^t + \alpha\exp{(-D)}\right)
\end{align}
with $\alpha \in [0,1]$.

\subsection{Efficient ProMises}
As explained in \cite{Andreella2022}, the Procrustes problem is well-posed if the term inside of the SVD is full-rank. However, in the case where $n$ the number of vertices exceeds $p$ the number of samples, the functional term $(F^s)^T F^t$ is not-full rank. To circumvent this, \cite{Andreella2022} proposes the Efficient Promises Model: functional data is first decomposed as $U^s, \Sigma^s, (V^s)^T = \svd\left(F^s\right)$ and $U^t, \Sigma^t, (V^t)^T = \svd\left(F^t\right)$. The inside term is now expressed as:
\begin{align}
    A = (1-\alpha) \Sigma^s  (U^s)^T  U^t  \Sigma^t + \alpha (V^s)^T  \exp(-D)  V^t
\end{align}
Since $\Sigma^s  (U^s)^T  U^t  \Sigma^t$ and $(V^s)^T  \exp(-D)  V^t$ are both full rank matrices of rank $p$, one can solve a reduced Procrustes problem in the space of dimension $p$:
\begin{align}
    U^r, \Sigma^r, (V^r)^T &= \svd\left(A\right) \\
    R^r &= U^r  (V^r)^T
\end{align}
The final mapping between vertices is then recovered by computing $R= V^s R^r  (V^t)^T$.

Notice the final mapping also has rank $p$. Thus, when discrepancy between the number of vertices and the number of samples is large, the transformation strongly compresses information. This observation includes the edge case $\alpha=1$ where the full-rank anatomical prior is also compressed to the $p$ by $p$ matrix $(V^s)^T  \exp(-D)  V^t$.

\begin{figure*}[!ht]
  \begin{center}
    \includegraphics[width=\textwidth]{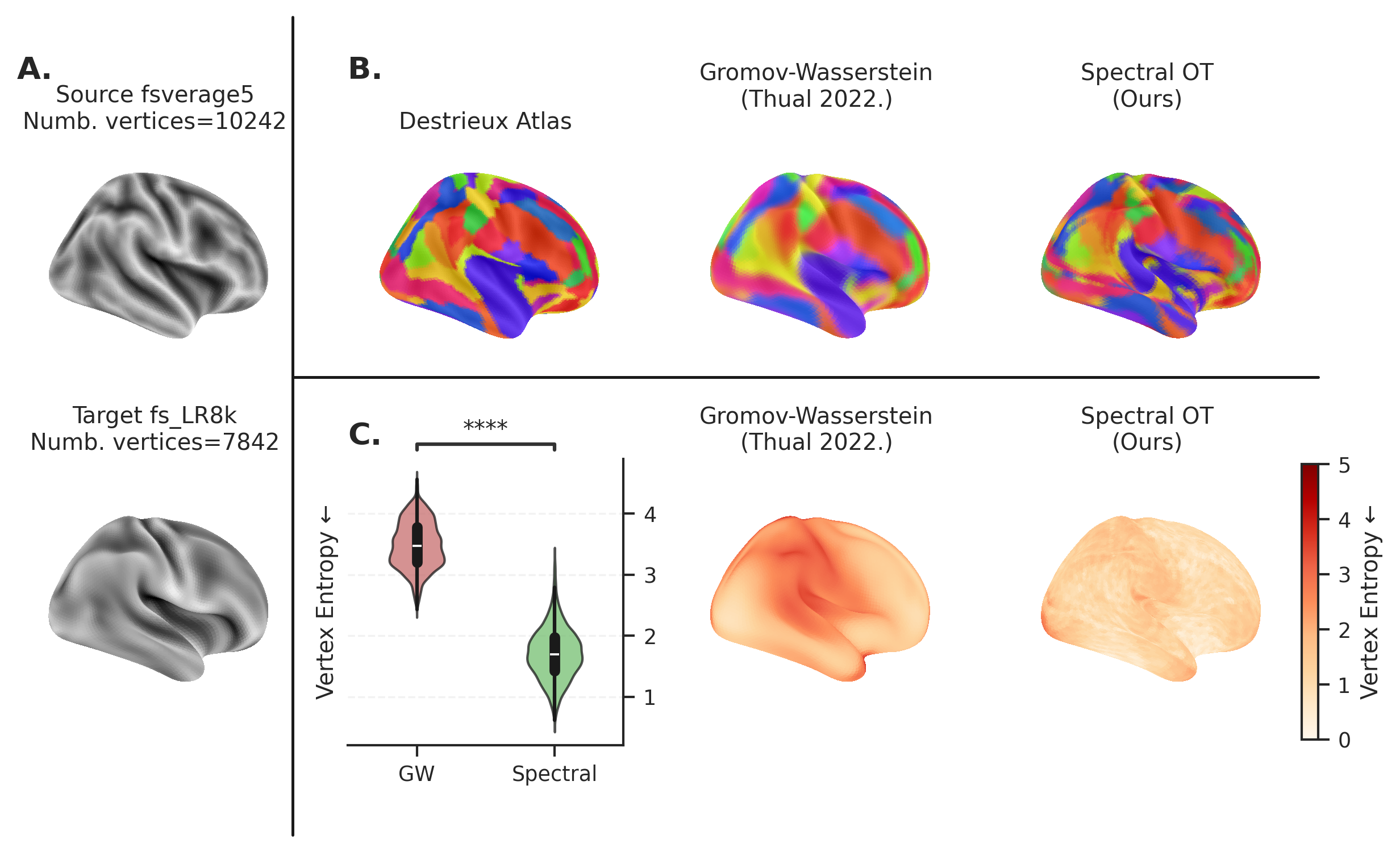}
  \end{center}
  \caption{\EDIT{
    \textbf{Anatomical alignment across different meshes ($\alpha=1$, right hemispheres).}
    Replication of experiment~1 on the right hemisphere.
    We set both OT solvers entropic regularization $\varepsilon$ to $10^{-4}$, with 1000 iterations of Sinkhorn's algorithm and 10 steps of block-coordinate descent for GW.
    \textbf{A.} We use right hemispheres of \emph{fsaverage5} and \emph{fsLR} surface templates as source and target respectively.
    \textbf{B.} Color transfer of the \emph{Destrieux} atlas between meshes using the computed transformations across vertices. Color blending is performed in RGB space to showcase information mixing across distinct areas of the atlas.
    \textbf{C.} Entropy of each target vertex probability distribution obtained after normalizing the transport plan across rows (lower is better).
  }}
  \label{fig:anatomical-example-appendix}
\end{figure*}

\begin{figure*}[!ht]
  \begin{center}
    \includegraphics[width=\textwidth]{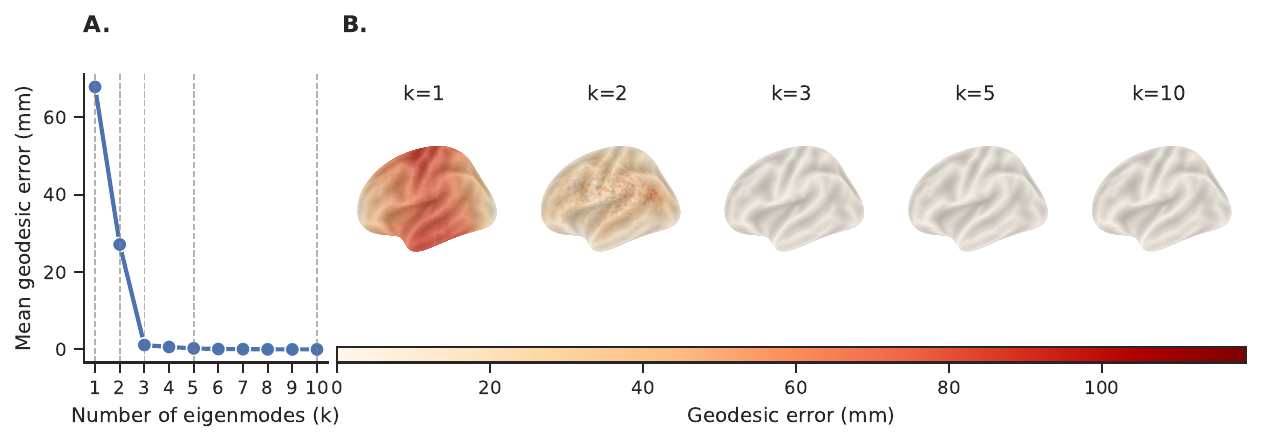}
  \end{center}
    \caption{\EDIT{
        \textbf{Influence of the number of eigenmodes on geodesic error.}
        We evaluate alignment accuracy using the \emph{geodesic error} metric, defined as the
        \emph{vertex displacement} from~\cite{thual2022aligning}: for each target vertex $j$,
        the displacement is computed as $\sum_i \mathbf{P}_{i,j}^{s\rightarrow t} \mathbf{D}_{i,j}
        / \sum_i \mathbf{P}_{i,j}^{s\rightarrow t}$, where $\mathbf{D}$ is the geodesic distance
        matrix on the mesh.
        Geodesic error is obtained by targeting the identity transformation (i.e., using the same
        mesh as source and target), so that a perfect alignment yields zero displacement.
        We use the left \emph{fsaverage5} pial surface as both source and target, with a purely
        anatomical setting ($\alpha=1$, $\varepsilon=10^{-3}$), and vary the number of eigenmodes
        $k$ used in the alignment.
        \textbf{A.} Mean geodesic error (mm) across vertices as a function of the number of
        eigenmodes $k$ (lower is better).
        \textbf{B.} Vertex-wise geodesic error projected onto the inflated mesh for selected values
        of $k$, indicated by the dotted vertical lines in \textbf{A}.
    }
    }
  \label{fig:eigenmodes-appendix}
\end{figure*}

\begin{figure*}[!ht]
  \begin{center}
    \includegraphics[width=.7\textwidth]{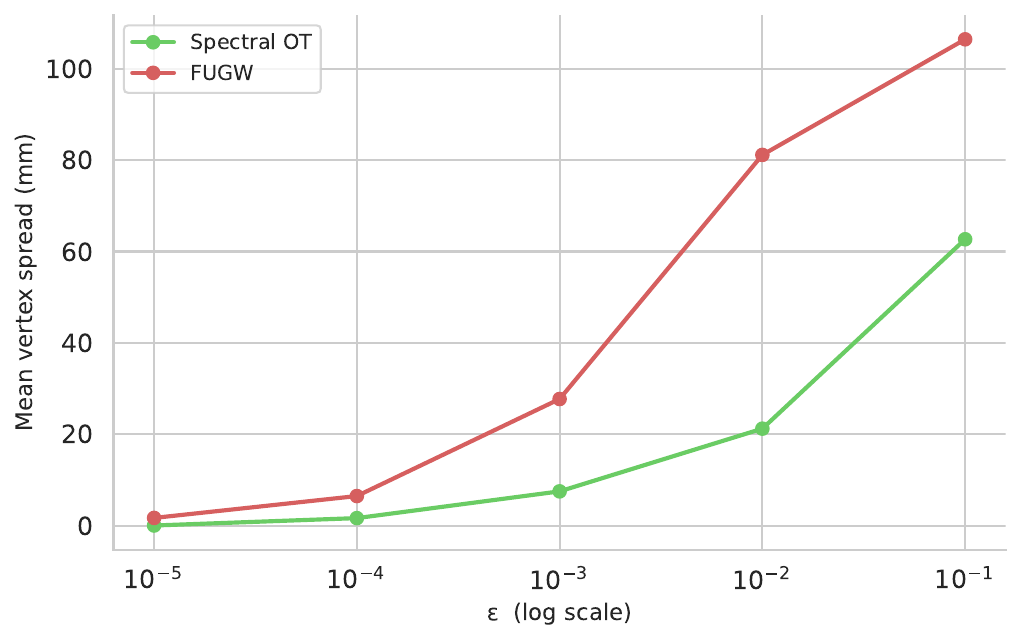}
    \caption{\EDIT{
        \textbf{Effects of entropic regularization on transformation blurriness.}
        We evaluate transformation blurriness using the \emph{vertex spread} criterion
        from~\cite{thual2022aligning}: for each source vertex $i$, $q$ pairs of indices
        $(j_q, k_q)$ are sampled from the row-normalized transport plan $\tilde{\mathbf{P}}_i$,
        and the spread is computed as the average pairwise geodesic distance
        $\frac{1}{q} \sum_{j_q, k_q} \mathbf{D}_{j_q, k_q}$ (lower is better).
        We recover the identity transformation on the left \emph{fsaverage5} mesh using
        $k=3$ anatomical eigenmodes ($\alpha=1$), and report the mean vertex spread across
        vertices as a function of entropic regularization $\varepsilon \in [10^{-5}, 10^{-1}]$,
        for both FUGW and SpectralOT.
    }}
  \label{fig:epsilon-appendix}
  \end{center}
\end{figure*}

\begin{figure*}[!ht]
  \begin{center}
    \includegraphics[width=.7\textwidth]{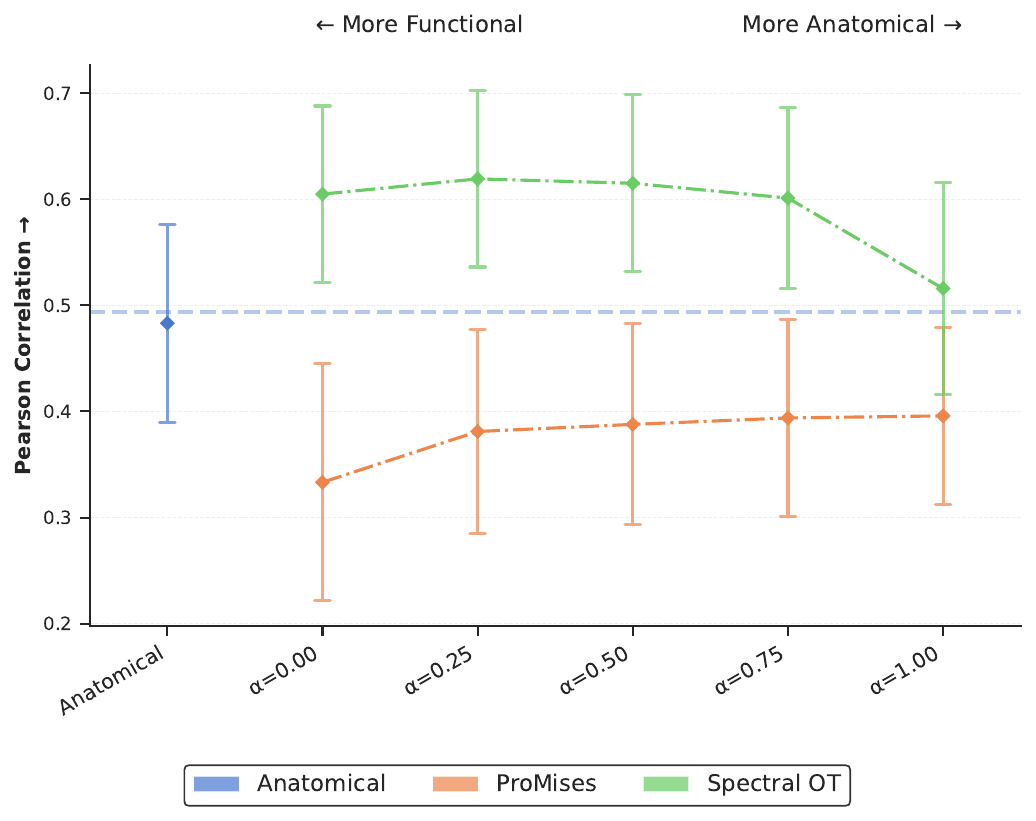}
  \end{center}
    \caption{\EDIT{
        \textbf{Inter-subject functional alignment on task data.}
        We replicate experiment~2 on all pairs of subjects from the \emph{MathLanguage} task,
        using paired samples from the \emph{Archi} task battery to derive the alignment
        transformation.
        Left-out \emph{MathLanguage} test data are mapped from each source subject to each
        target subject, and alignment quality is assessed by the Pearson correlation between
        aligned source and target test samples, averaged across all subject pairs.
        We compare SpectralOT ($\varepsilon = 10^{-3}$) against the ProMises model across
        varying levels of anatomical regularization. FUGW is excluded due to prohibitive
        computation time over all subject pairs.
        Anatomical registration on both \emph{fsaverage5} hemispheres serves as the baseline.
    }
    }
  \label{fig:isc-averaged-appendix}
\end{figure*}

\begin{figure*}[!ht]
  \begin{center}
    \includegraphics[width=.7\textwidth]{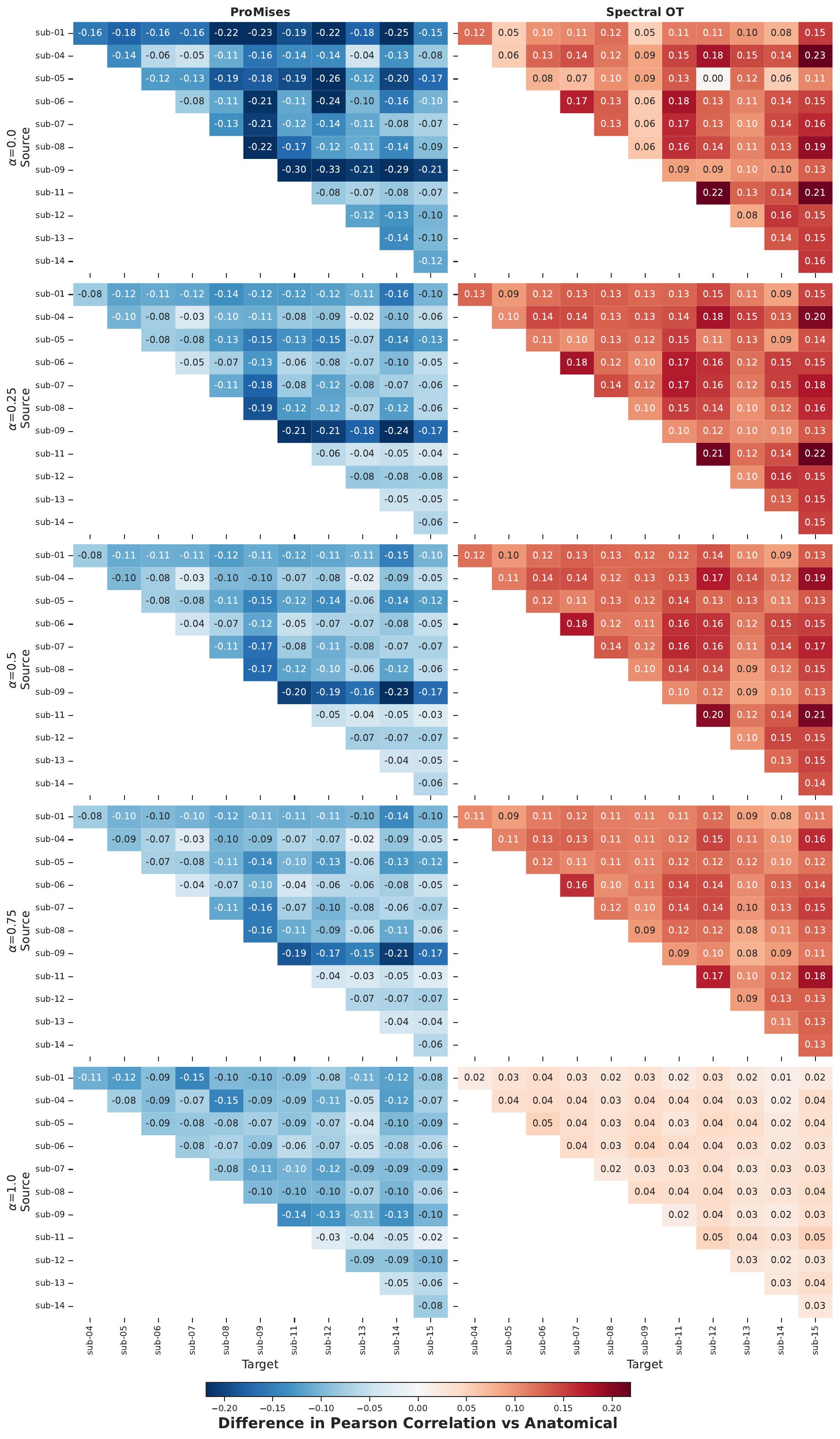}
    \caption{\EDIT{
        \textbf{Detailed inter-subject correlations for experiment~2.}
        Difference in Pearson correlation relative to the anatomical baseline, shown for
        every increasing source/target subject pair in experiment~2, for both ProMises and SpectralOT.
        Each entry is averaged across pairs of samples, and warmer colors indicate a larger
        improvement over baseline.
    }
    }
  \label{fig:isc-complete-appendix}
  \end{center}
\end{figure*}

\end{document}